\documentstyle[prl,aps,multicol]{revtex}

\begin{document}

\title{No-cloning of Orthogonal States in Composite Systems}

\author{Tal Mor\thanks{DIRO, Universit\'e de Montr\'eal, Montr\`eal, Canada}}

\date{12/Feb/97}

\maketitle

\begin{abstract}

The no-cloning principle tells us that non-orthogonal quantum states
cannot be cloned, but it does not tell us that orthogonal states can 
always be cloned. 
We suggest a situation where the cloning transformations are restricted, 
leading to a novel type of no-cloning principle.
In the case of a composite system made of two subsystems: if the 
subsystems are only available one after the other then there
are various cases when {\em orthogonal states} cannot be cloned. 
Surprising examples are given, which give a radically better insight
regarding the basic concepts of quantum cryptography. 
\end{abstract}

\begin{multicols}{2}

The no-cloning theorem describes one of   
the most fundamental non-classical properties of quantum systems.
It states that an unknown quantum state cannot 
be cloned~\cite{WZ82,Dieks82}.
Assuming that such a quantum pure state $\rho$ 
can be cloned leads to a violation of
unitarity of quantum mechanics.
Even if there is some information about the state (e.g., it is either
$\rho_0$ or $\rho_1$) it cannot be cloned in the general case.
This means that we cannot create a copying device which gets the unknown
state ($\rho_p$) as an input and produces two copies of it
at the output.
Another version of the no-cloning theorem~\cite{BBM} states that 
any attempt of learning something regarding 
the input state of the copying device
(even an attempt of making a very faint imprint) will necessarily
induce some disturbance in the output state;
This principle presents a very interesting variation of the 
so called ``uncertainty principle'' since it applies to an individual
system;
see~\cite{FP} for more details.

Let 
$\rho_0 \equiv |\phi_0\rangle \langle \phi_0|$  
and $\rho_1 \equiv |\phi_1\rangle \langle \phi_1|$  
be two non identical pure states 
provided by the producer (Alice),
and suppose that these states are known to the person (Eve) who attempts 
to clone them.
Let Eve receive one of them ($\rho_p$) ---
without knowing which one ---
as an input of her cloning (copying) device,
and assume the initial normalized pure state of the cloning device is $E$. 
Then, a success to clone, that is, to create a product state 
$\rho_p \rho_p$
from the unknown input state $\rho_p$,
is described by 
\begin{equation}
E \rho_p \longrightarrow {E_p}' \rho_p \rho_p
\end{equation}
where the dimensionality of the primed system $E'$ 
is smaller than the dimensionality of $E$.
This process violates unitarity
if the states are non-orthogonal: unitarity promises that 
$ {\rm Tr} (EE){\rm Tr}(\rho_0 \rho_1) = 
 {\rm Tr}({E_0}' { E_1}') 
 {\rm Tr}(\rho_0 \rho_1)
 {\rm Tr}(\rho_0 \rho_1) $;
using the normalization conditions we get
 ${\rm Tr} (E E) = 1$ and
 ${\rm Tr} ({E_0}' {E_1}') \le 1$,  
so for non-identical states with ${\rm Tr} (\rho_0 \rho_1) < 1$, 
cloning is impossible unless 
${\rm Tr}(\rho_0 \rho_1) = 0$. 
Hence, non-identical states can be cloned only if they are orthogonal.
In the cloning process above, Eve knows the set
of states $\rho_p$, and therefore, in case they are identical she can
clone them;
the cloning device creates the state
$\rho=\rho_0=\rho_1$ without measuring anything.
In the more effective version of
the no cloning theorem~\cite{BBM} the states changes according to
\begin{equation}
E \rho_p \longrightarrow E_p \rho_p \ .
\end{equation}
In this case unitarity promises us that 
$ {\rm Tr} (E E)    
  {\rm Tr} (\rho_0 \rho_1) = 
 {\rm Tr}(\rho_0 \rho_1){\rm Tr}(E_0 E_1) $,
leading to ${\rm Tr}(E_0 E_1) = 1$ for (non-identical) non-orthogonal states.
Thus $E_0$ and $E_1$ are identical and can provide
no information on $\rho_p$.
We shall refer to this process as {\em no-imprint principle} to distinguish it
from the {\em no-cloning principle} of Eq.~(1).
If $E_0$ and $E_1$ are not identical (hence, 
provide information) in an imprint process, 
then the output states cannot be identical to the input states $\rho_p$.
See~\cite{FP} for detailed analysis of information vs.~disturbance in case
of non-zero disturbance in an imprint process.

These conclusions apply also when $\rho_p$ are mixed states,
telling us that only identical or orthogonal states $\rho_p$ can be cloned
using the first process~\cite{BCFJS96} or imprinted using the second process.
However, the situation is more delicate since 
such processes do not provide a complete 
description of Eve's possible strategies when $\rho_p$ are mixed states!
The no-cloning of quantum mixed states was recently 
analyzed, and it was shown that, 
while commuting non-orthogonal quantum mixed states cannot be cloned
[using Eq.~(1)], 
they can still be 
{\em broadcast}~\cite{BCFJS96}: 
Eve can create a state $\chi_p$ such that 
\begin{equation} \chi_p \ne \rho_p \otimes \rho_p
\end{equation}  but which satisfies 
\begin{equation}
{\em Tr}_E[\chi_p] = {\em Tr}_A[\chi_p] = \rho_p
\ .
\end{equation}
A similar extension of Eq.(2) to achieve a ``broadcast-imprint''
process is straight forward. Surprisingly, it
is not restricted to commuting density matrices~\cite{Fuchs},
and a complete analysis is still missing.
Due to Eq.~(3) (or its counterpart when performing a
broadcast-imprint process of mixed states),
the state of the cloning device is entangled
with the state of the system.

These no-cloning theorems prove that non-orthogonal states cannot be cloned, 
and that 
cloning of orthogonal states is possible if arbitrary unitary transformation
can be chosen.
In this work we suggest to {\em restrict} the allowed unitary 
transformations.
We show that, as result of this restriction, 
there are orthogonal states which {\em cannot
be cloned}.
We suggest a particular restriction, 
where $\rho_0$ and $\rho_1$ are two orthogonal states of a composite system
and the subsystems from which the system is composed are only available
(to the cloning device) one after the other.
Other restrictions are also possible, and lead to very fascinating
examples~\cite{Shor}.

The restriction of the type we use here is typical in quantum key distribution.
We show that ``no cloning of orthogonal states'' is the basic principle used
in many quantum key distribution schemes, 
rather than the standard no-cloning arguments,
as was previously argued and believed.
Hence, we shed new light on the possible basic concepts 
which are at the roots of 
secure quantum key distribution.

Let the system $A$ in Alice's hands be composed of two subsystems 
$A_1$ and $A_2$, such that the possible states provided by Alice,
$\Phi_p(A_1 A_2)$, are orthogonal to each other, in the Hilbert space of
the composite system.
If the two subsystems were provided to Eve together 
she could clone the states.
However, the subsystems are provided to Eve only one after the other
so she cannot access the second one while she holds the first.
Let $\rho_p(A_1) = {\rm Tr}_{A_2} (\Phi_p)$ be the reduced density matrices
of the first subsystem, and assume that the two (or more)
possible states are non-identical and non-orthogonal.
Consider Eve's possible strategies when she holds the first subsystem $A_1$.
Clearly, if she changes $\rho_p(A_1)$ before letting it go her cloning 
attempt fails since she will have no access to that subsystem in the future.
Thus, she cannot use the cloning process or imprint process.
However, she can still use a broadcast-cloning process or a
broadcast-imprint process.
We only need to verify that these 
processes shall not allow her to clone in our scenario:
Indeed, if the states of the first subsystem are non-orthogonal,
Eve might be able to achieve a broadcast, 
but the state of the first subsystem necessarily becomes 
entangled
with Eve's state, hence cannot be fully entangled 
(or fully correlated) 
with the second subsystem anymore.
Therefore, although the state $\rho_p(A_1)$ does not change in the broadcast 
process, the state $\Phi_p$ necessarily changes, and noise is induced.
The last thing we should worry about is that Eve will not be able to clone by
obtaining the entire information from the second subsystem, meaning that 
$\rho_p(A_2) = {\rm Tr}_{A_1}(\Phi_p)$ must be non orthogonal as well.

Thus, based on the previous discussion, we reach a novel 
{\em no cloning principle for orthogonal states.} --- 
The two (or more) orthogonal states $\rho_p(A_1 A_2)$ 
of the system composed of $(A_1)$ and $(A_2)$ cannot be 
cloned if the reduced density matrices 
of the subsystem which is available first (say $A_1$) 
\begin{equation} 
\rho_p(A_1) = {\rm Tr}_{A_2} [\rho_p(A_1 A_2)]
\end{equation}  are non orthogonal 
and non identical, and if the reduced density matrices of the second
subsystem are non orthogonal.

The first case we study is the case where the two subsystems are entangled,
such that the first subsystem $A_1$
is in one of two commuting mixed states.
Let $A_1$ and $A_2$ be two qubits with $| 0_b \rangle$ and $| 1_b \rangle$ the 
basis vectors of the $b$'th qubit.
Let the initial states be the two orthogonal states~\cite{Imoto97} 
\begin{equation} \psi_0 = \cos \alpha | 0_1 \otimes 1_2 \rangle 
                             + \sin \alpha | 1_1 \otimes 0_2 \rangle 
\end{equation}
\begin{equation} \psi_1 = \sin \alpha | 0_1 \otimes 1_2 \rangle 
                             - \cos \alpha | 1_1 \otimes 0_2 \rangle 
\ ,
\end{equation}
and $\alpha$ is known to the cloner.
In case $\alpha = 0$ or $\alpha = \pi/2$ all data is already contained 
in the first particle. In terms of cloning of the first subsystem -- 
the two possible reduced states are orthogonal, hence can be cloned.
In case of $\alpha = \pi/4$ (so that $\cos \alpha = \sin \alpha = 1/\sqrt 2$)
the two possible reduced 
states of the first subsystem
are identical.
Thus, Eve can release a dummy qubit (denoted by $E_3$)
of her own,
entangled with another one (denoted by $E_4$), say in a state
$ \psi_0 = (1/\sqrt2) | 0_3 \otimes 1_4 \rangle 
                             + (1/\sqrt2)  | 1_3 \otimes 0_4 \rangle $,
while keeping subsystem $A_1$. 
The reduced state of the dummy particle is equal to the state of the
first subsystem.
Later on, after receiving subsystem
$A_2$ and learning the state of the combined system
$A_1 A_2$, Eve can change the state of ($E_3 E_4$)
by a simple transformation on particle $E_4$ alone. 
In terms of cloning of the first system -- 
the two possible states are the same,
the completely mixed state, hence can be cloned.
For all other $\alpha$, there is no strategy for Eve to learn the data,
or even to get some information unless disturbing the state.
The reduced density matrix of the first particle (particle $1$) are
\begin{equation}
\rho_0 =     \cos^2 \alpha | 0 \rangle \langle 0 |
          +  \sin^2 \alpha | 1 \rangle \langle 1 |
\end{equation}
and
\begin{equation}
\rho_1 =     \sin^2 \alpha | 0 \rangle \langle 0 |
          +  \cos^2 \alpha | 1 \rangle \langle 1 |
\ .
\end{equation}
The two density matrix commutes, but they are not orthogonal nor identical.
Therefore they can be broadcast, but they can not be cloned,
meaning that the resultant state of of $A_1$ and $E$ can be 
$\chi_p(A_1 C)$ such that 
${\rm Tr}_E[\chi_p(A_1 E)] = \rho_p$ and 
${\rm Tr}_{A_1}[\chi_p(A_1 E)] = \rho_p$ but 
$\chi_p(A_1 C)$ is not a tensor product of these two matrices.
Therefore, Eve's system is entangled with $A_1$, and some noise is 
necessarily introduced since the resulting (three-particle) state is
not a tensor product of Alice's (two-particle) state and Eve's state.

This example (in a somewhat different form which uses Fock states) 
was suggested 
recently by Koashi and Imoto~\cite{Imoto97} for quantum key distribution.
It was suggested as a generalization and modification of
a previous scheme of Goldenberg and Vaidman~\cite{GV95}.
Goldenberg and Vaidman (GV) were the first to realize that
one can use orthogonal states for quantum key distribution.
Their work~\cite{GV95} emphasizes that 
non-orthogonal states are not crucial in quantum cryptography,
but it does not discuss it in terms of no-cloning of the reduced
density matrices of the first subsystem as we do here.
GV`s scheme is using random timing, 
since it uses the case
of $\alpha = \pi/4$ which, without random timing, is insecure. 
Hence, it actually uses (as is explained in~\cite{Imoto97})
three orthogonal states,  where the third one is 
\begin{equation} \psi_2 = |0_1\rangle |0_2\rangle \ ,
\end{equation}
(which is the vacuum state in their description) with a reduced density matrix 
\begin{equation}
\rho_2 = | 0 \rangle \langle 0 | \ .
\end{equation}
This state $\psi_2$ is used only for error verification.
[To observe that indeed~\cite{GV95} used three states, one must describe 
their work using Fock states $|0^{(a)} \rangle $ etc.~in arms $a$ and $b$]. 
Note that in this case, the three possible reduced density matrices of the 
first particle are still commuting, but they are {\em not identical},
hence can be broadcast but not cloned.

In a comment to~\cite{GV95}, Peres~\cite{Peres96}
emphasizes that, as far as Alice is concerned, previous schemes 
for quantum cryptography such as 
the original (four-state) key distribution scheme of~\cite{BB84}
also use orthogonal states, since, in the four-state scheme
Alice chooses one of two orthogonal states
in a basis of her choice. Peres also emphasizes the need of a second stage,
where classical information is provided to Eve 
(regarding the basis in the four-state scheme or the random timing in
GV scheme)
when Eve has no longer access to the quantum data.
See also Goldenberg-Vaidman reply~\cite{GV96}. Note that we do not consider
the case of spatial separation, but only the case of ``time separation''.

Using our new form of the no cloning argument 
we provide a stronger statement:
in both the GV scheme and the four-state scheme  
the states which are {\em transmitted through Eve} are orthogonal!
This is a very surprising result, since (a).---it tells us that 
entanglement is not vital for preventing cloning of orthogonal states,
and since (b).---it was always believed that the four-state scheme uses four 
non-orthogonal states (see explanation in~\cite{GV96} for instance).

In the following we show striking similarities 
between GV scheme and the four-state scheme, which are far  
beyond the similarity shown by Peres.
To achieve this understanding we present a non-standard,
fully quantum description 
of the four-state scheme, which however, is completely equivalent
to the standard description.
We claim that the standard belief that quantum key distribution is based
upon the no-cloning theorem of Eq.~(2) is inappropriate:
In quantum key distribution Alice's states are sent to another person 
(Bob) who must be able to learn them (at least sometimes).
The states are sent through Eve.
It is not enough to prevent Eve from learning the encoded bits, but we also
need to make sure that Bob can learn them.
Thus, a standard no-cloning argument does not suffice as the basis 
of quantum key distribution.
Indeed, using  a complete quantum 
description, we show that  
our no-cloning argument for orthogonal states is actually used in the standard
four-state scheme of~\cite{BB84}.

Let $0$ and $1$ form a basis of two dimensional Hilbert space
and let $0^x = (1/\sqrt2)[0 + 1]$
and $1^x = (1/\sqrt2)[0 - 1]$.
Let Alice prepare one of the following four states
$|\phi_{0_z}\rangle = |0_1 0_2 \rangle $,
$|\phi_{0_x}\rangle = |0^x_1 1_2 \rangle $,
$|\phi_{1_z}\rangle = |1_1 0_2 \rangle $, and 
$|\phi_{1_x}\rangle = |1^x_1 1_2 \rangle $, 
which are all orthogonal to each other.
The first qubit (qubit $1$) is in one of four pure states
$|0\rangle$;
$|1\rangle$;
$(1/\sqrt2)|0+1\rangle$;
$(1/\sqrt2)|0-1\rangle$,
which are the states used in the conventional form of~\cite{BB84}, 
and the second qubit 
carries classical information telling whether the first qubit was prepared in
one basis or the other.

Consider the two mixed states
\begin{equation} \chi_0 = 
|\phi_{0_z}\rangle  \langle \phi_{0_z}| +
|\phi_{0_x}\rangle  \langle \phi_{0_x}|
                        = |0_1 0_2 \rangle \langle 0_1 0_2 | 
                          + |0^x_1 1_2 \rangle \langle 0^x_1 1_2 | 
\end{equation}
and
\begin{equation} \chi_1 = 
|\phi_{1_z}\rangle  \langle \phi_{1_z}| +
|\phi_{1_x}\rangle  \langle \phi_{1_x}|
                        = |1_1 0_2 \rangle \langle 1_1 0_2 | 
                          + |1^x_1 1_2 \rangle \langle 1^x_1 1_2 | 
\ ,
\end{equation}
which are used to encode $0$ and $1$ respectively,
and are sent to Bob through Eve.
[Note that combining each pair of states into one mixture is a result of
the fact that the secret bit (the data that should be cloned) in~\cite{BB84}
is defined by these mixed states].
The two mixed states $\chi_p$ are orthogonal, 
since each of the pure states which
decompose one of the matrices is orthogonal to the states which decompose
the other.
The reduced density matrices of particle $1$ (when particle $2$
is traced out from the states $\chi_p$) again have the form
\begin{equation}
\rho_0 =     \cos^2 \alpha | 0' \rangle \langle 0' |
          +  \sin^2 \alpha | 1' \rangle \langle 1' |
\end{equation}
and
\begin{equation}
\rho_1 =     \sin^2 \alpha | 0' \rangle \langle 0' |
          +  \cos^2 \alpha | 1' \rangle \langle 1' |
\ 
\end{equation}
(in a basis, denoted by the prime, which is exactly between
the standard ($z$) basis and the $x$ basis used above, and is also known 
as the Breidbart basis),
with $\cos^2 \alpha= (1+1/\sqrt2)/2$. 
These density matrices are commuting but not orthogonal.
Therefore the two density matrices
can be broadcast but not cloned, and any attempt
of cloning them will create noise.

Note that we ignored the irrelevant step in which Bob is telling
Alice his choice of basis.
This step is required in the original protocol due to technological 
limitations, and can be eliminated once Bob keeps the first particle 
(in a quantum state) till receiving the second.
However, it is crucial that Bob receives the first particle {\em before}
Alice sends
the second and communication from Bob to Alice is required to verify this.

This far we have seen the impossibility to clone three orthogonal states 
when two are entangled~\cite{GV95}, the impossibility to clone
two orthogonal entangled states~\cite{Imoto97} and the impossibility to
clone four non-entangled (product) 
states~\cite{Peres96} or two non-entangled (sum of product)
density matrices (our description of~\cite{BB84}).
Let us search for simpler cases, with non entangled orthogonal states.
Such a case will make the best use of the new no-cloning theorem stated
in this work. Furthermore, 
such orthogonal states which cannot be cloned are useful for quantum key 
distribution, and might be simpler to implement, or to analyze,
relative to the other schemes. 

It is impossible to use only two product pure states:
Let $|\phi_0\rangle = |{0_1}' {0_2}'' \rangle$ 
(which is general since the basis
is arbitrary), and $|\phi_1 \rangle $ a state which is orthogonal to it;
Clearly,  
$|\phi_1\rangle = |{1_1}' \chi_2 \rangle$ or 
$|\phi_1\rangle = | \chi_1  {1_2}'' \rangle$ 
(with arbitrary state $\chi$, and with $1$ orthogonal 
to $0$ in each primed or double-primed basis)  
are the only two possibilities to choose a two-particle
state orthogonal to $|\phi_0\rangle$. For any of these choices the entire
information can be cloned by Eve by cloning the appropriate particle
(e.g. the first particle in the first choice of $|\phi_1\rangle$);
similar argument is true when using higher dimensions or larger 
number of particles.

Note that we did not describe the scheme of~\cite{Ben92} in terms of 
cloning, since a more complicated involvement of the 
third party (Bob) is needed for its description.
Two-way communication between Alice and Bob
plays a vital role in this scheme,
and it is not clear yet if our new 
no-cloning principle suffices to explain why this scheme works.  

The natural simplest (achievable) possibilities are 
to use only three orthogonal pure states or two pure 
orthogonal states and one mixed state which is orthogonal to both;
as far as quantum key distribution is concerned, let two pure non entangled
states carry the secret key, and the third non entangled
state (pure or mixed) provide the protection from cloning (this is called 
``data rejected protocol''):
The states 
\begin{equation} | \phi_0 \rangle =|0_1 0_2 \rangle \ ;
\quad | \phi_1 \rangle =|1_1 0_2 \rangle \end{equation}  
will be two states
which carry the key (in the first qubit), and in addition the state 
\begin{equation} 
| \phi_2 \rangle  = |0^{(x)}_1 1_2 \rangle \ ,  \end{equation}  
or alternatively, an equal mixture of this state and 
$ |1^{(x)}_1 1_2 \rangle $,
\begin{equation} 
 \chi_2 =        |0^{(x)}_1 1_2 \rangle  
                                         \langle 0^{(x)}_1 1_2 | + 
                 |1^{(x)}_1 1_2 \rangle  
                                         \langle 1^{(x)}_1 1_2 | 
\ ,\end{equation}
provide the protection from cloning. 
Such schemes (especially the second)
are as simple as~\cite{BB84} as far as practical implementation
is concerned, but might be much simpler to analyze, since Eve can only
obtain information in one basis.

In this work we discussed the impossibility to clone orthogonal
states of combined systems in various case, based on the impossibility
to clone non orthogonal mixed states. 
We presented a unified formalism for 
several schemes in quantum key distribution based on our new no-cloning
argument for orthogonal states.
We corrected
an unjustified claim regarding the role of Eq.~(2) in quantum key distribution,
and solved a dispute regarding
the use of orthogonal states in quantum key distribution.
Finally we suggested the simplest schemes which are based on this principle.

I would like to thank Gilles Brassard, Chris Fuchs, Lior Goldenberg
and Asher Peres for very helpful discussions.
This work was supported in part by Canada's {\sc nserc} and Qu\'ebec's
{\sc fcar}.
Part of this work was completed during the 1997 Elsag-Bailey --- I.S.I.
Foundation research meeting on quantum computation.

\end{multicols}
\end{document}